\documentclass[runningheads]{llncs}
\usepackage{graphicx}
\usepackage{bm}
\usepackage{amssymb,amsmath}
\usepackage{cite}
\usepackage{color}
\usepackage{soul}

\begin{document}
\title{Learning Hidden States in a Chaotic System:\\A Physics-Informed Echo State Network Approach\thanks{The authors acknowledge the support of the Technical University of Munich - Institute for Advanced Study, funded by the German Excellence Initiative and the European Union Seventh Framework Programme under grant agreement no. 291763. L.M. also  acknowledges the Royal Academy of Engineering Research Fellowship Scheme.}}
\titlerunning{Leaning Hidden States in a Chaotic System: a PI-ESN Approach}
%
\author{N.A.K. Doan\inst{1,2} \and
W. Polifke\inst{1} \and
L. Magri\inst{3,4}
}
\authorrunning{N.A.K. Doan et al.}
%
\institute{Department of Mechanical Engineering, Technical University of Munich, Germany \and
Institute for Advanced Study, Technical University of Munich, Germany \and
Department of Engineering, University of Cambridge, United Kingdom \and
(visiting) Institute for Advanced Study, Technical University of Munich, Germany}
\maketitle              
\begin{abstract}
We extend the Physics-Informed Echo State Network (PI-ESN) framework to reconstruct the evolution of an unmeasured state (hidden state) in a chaotic system. 
The PI-ESN is trained by using (i) data, which contains no information on the unmeasured state, and (ii) the physical equations of a prototypical chaotic dynamical system. 
Non-noisy and noisy datasets are considered.
First, it is shown that the PI-ESN can accurately reconstruct the unmeasured state.
Second, the reconstruction is shown to be robust with respect to noisy data, which means that the PI-ESN acts as a denoiser. 
This paper opens up new possibilities for leveraging the synergy between physical knowledge and machine learning to enhance the reconstruction and prediction of unmeasured states in chaotic dynamical systems.
%

\keywords{Echo State Networks \and Physics-Informed Echo State Networks \and Chaotic Dynamical Systems \and State Reconstruction.}
\end{abstract}

\section{Introduction}
In physical science experiments, it is often difficult to measure all the physical states, whether it be because the instruments have a finite resolution, or because the measurement techniques have some limitations. Consequently, we are typically able to infer only a few states of the system from the measured observable quantities. The states that cannot be measured are {\it hidden}, that is, they may affect the system's evolution, but they cannot be straightforwardly measured.
The accurate reconstruction of hidden states is crucial in many fields such as cardiac blood flow modelling \cite{Sankaran2012}, climate science \cite{Kalnay2003}, and fluid dynamics \cite{Brenner2019}, to name only a few. For example, in fluid dynamics, measurements of the velocity field with particle image velocimetry may be limited to the in-plane two-dimensional velocity, although the three-dimensional velocity is the quantity of interest. 
The reconstruction of unmeasured quantities from experimental measurements has been the subject of recent studies, that used a variety of data assimilation and/or machine learning techniques. For example, spectral nudging, which combines data assimilation with physical equations, was used to infer temperature and rotation rate in 3D isotropic rotating turbulence \cite{ClarkDiLeoni2018}. Alternatively, \cite{Fukami2018} reconstructed the fine-scale features of an unsteady flow from large scale information by using a series of Convolutional Neural Networks. Using a similar approach, the reconstruction of the velocity from hydroxyl-radical planar laser induced fluorescence images in a turbulent flame was also performed \cite{Barwey2019}. Another approach based on echo state networks has also been used for the reconstruction of time series of unmeasured states of chaotic systems \cite{Lu2017}. While effective in reconstructing the unmeasured states, these approaches required training data with both the measured {\it and} unmeasured states.
In this paper, we propose using physical knowledge to reconstruct hidden states in a chaotic system \textit{without the need of any data of the unmeasured states during the training}. This is performed with the Physics-Informed Echo State Network (PI-ESN), which has been shown to be suited to the accurate forecasting of chaotic systems \cite{Doan2019}.
The PI-ESN, and more generally \textit{Physics-Informed} Machine Learning, relies on using the physical knowledge of the system under study, in the form of its conservation equations, and adding the computations of these conservation equations in the loss function during the training of the machine learning framework \cite{Raissi2018b,Doan2019}. These approaches, which combine physical knowledge and machine learning, have been shown to be efficient in improving the accuracy of neural networks \cite{Raissi2018b,Doan2019}.
The PI-ESN approach will be, here, applied to the Lorenz system, which is a prototypical chaotic system~\cite{Lorenz1963}.  

The paper is organized as follows. The problem statement and the methodology based on PI-ESN are detailed in Sect. \ref{sec:method}. Then, results are presented and discussed in Sect. \ref{sec:results} and final comments are summarized in Sect. \ref{sec:conclusion}.

\section{Methodology: Physics-Informed Echo State Network for learning of hidden states}
\label{sec:method}
We consider a dynamical system whose governing equations are:
\begin{equation}
\mathcal{F}(\bm{y}) \equiv \dot{\bm{y}} + \mathcal{N} (\bm{y}) = 0
\label{eq:dynamic}
\end{equation}
where $\mathcal{F}$ is a non-linear operator, $\dot{~}$ is the time derivative and $\mathcal{N}$ is a nonlinear differential operator. Eq. (\ref{eq:dynamic}) represents a formal ordinary differential equation, which governs the dynamics of a nonlinear system. It is assumed that only a subset of the system states can be observed, which is denoted $\bm{z} \in \mathbb{R}^{N_z}$, while the hidden states are denoted $\bm{h} \in \mathbb{R}^{N_h}$. The full state vector is $\bm{y} \in \mathbb{R}^{N_y}$, which is the concatenation of $\bm{z}$ and $\bm{h}$, i.e., $\bm{y} = [\bm{z};\bm{h}]$. The vectors' dimensions are related by $N_y = N_z + N_h$. The objective is to train a PI-ESN to reconstruct the hidden states, $\bm{h}$. We assume that we have training data of the measured states $\bm{z}(n)$ only, where $n=0,1,2,\ldots, N_t-1$ are the discrete time instants that span from $0$ to $T=(N_t-1)\Delta t$, where $\Delta t$ is the sampling time. Thus, the specific goal for the PI-ESN is to reconstruct the hidden time series, $\bm{h}(n)$, for the same time instants. To solve this problem, the PI-ESN of \cite{Doan2019}, which is based on the \textit{data-only} ESN of \cite{Lukosevicius2009}, needs to be extended, as explained next.

\begin{figure}[!ht]
	\centering
	\includegraphics[width=0.49\textwidth]{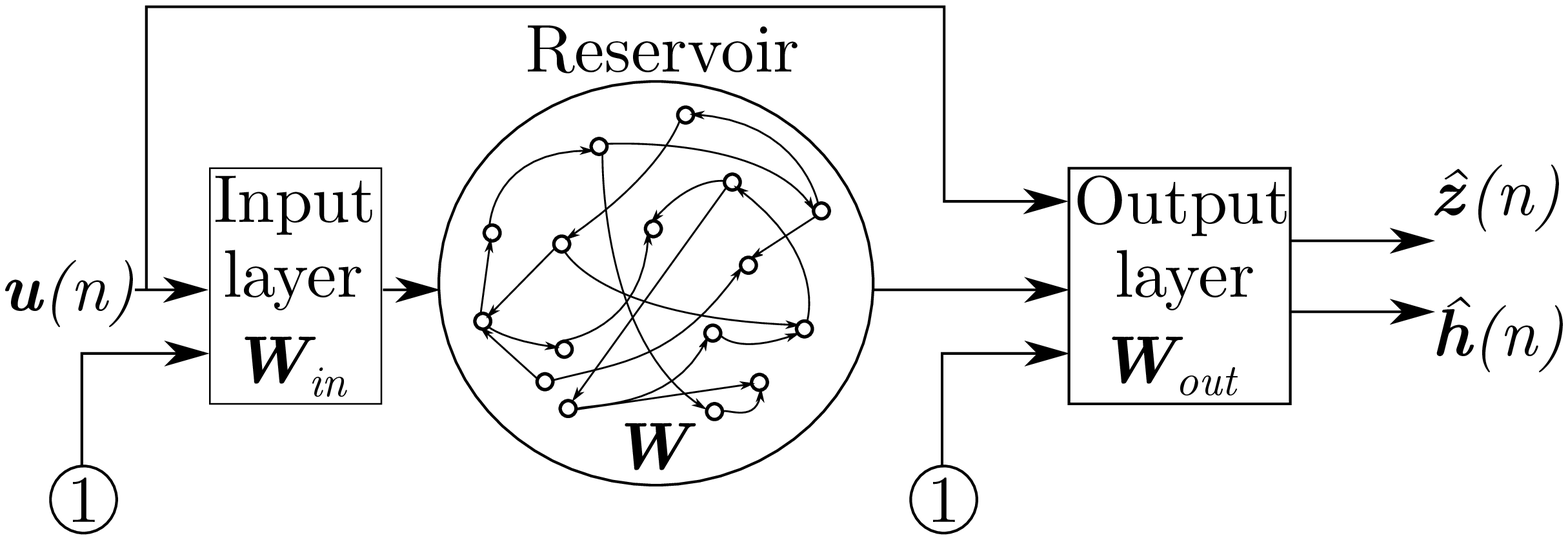}
	\caption{Schematic of the ESN. 
	\raisebox{.5pt}{\textcircled{\raisebox{-.9pt} {1}}} indicates the bias.
	}
	\label{fig:ESN_schema}
\end{figure}

The PI-ESN is composed of three main parts (Fig. \ref{fig:ESN_schema}): 
(i) an artificial high dimensional dynamical system, i.e., the reservoir, whose neurons' (or units') states at time $n$ are represented by a vector, $\bm{x}(n) \in \mathbb{R}^{N_x}$, representing the reservoir neuron activations; 
(ii) an input matrix, $\bm{W}_{in}\in \mathbb{R}^{N_x \times (1+N_u)}$, and (iii) an output matrix, $\bm{W}_{out}\in \mathbb{R}^{N_y \times (N_x+N_u+1)}$. The reservoir is coupled to the input signal, $\bm{u}\in \mathbb{R}^{N_u}$, via $\bm{W}_{in}$. A bias term is added to the input to excite the reservoir with a constant signal. The output of the PI-ESN, $\widehat{\bm{y}}$, is a linear combination of the reservoir states, inputs and an additional bias:
\begin{equation}
    \label{eq:ESN_output}
    \widehat{\bm{y}}(n)=[\widehat{\bm{z}}(n); \widehat{\bm{h}}(n)] = \bm{W}_{out} [\bm{x}(n);\bm{u}(n);1]
\end{equation}
where $[;]$ indicates a vertical concatenation and $\widehat{\cdot}$ denotes the predictions from the PI-ESN. The PI-ESN outputs both the measured states, $\widehat{\bm{z}}$, and the hidden states, $\widehat{\bm{h}}$ (Eq. (\ref{eq:ESN_output})). The reservoir states evolve as:
\begin{equation}
\bm{x}(n) = \tanh \left( \bm{W}_{in} [\bm{u}(n);1] + \bm{W} \bm{x}(n-1) \right)
\end{equation}
where $\bm{W}\in\mathbb{R}^{N_x\times N_x}$ is the recurrent weight matrix and the (element-wise) $\tanh$ function is the activation function for the reservoir neurons.
Because we wish to predict a dynamical system, the input data for the PI-ESN corresponds to the measured system state at the previous time instant, $\bm{u}(n) = \bm{z}(n-1)$, which is only a subset of the state vector. In the ESN approach \cite{Lukosevicius2009}, the input and recurrent matrices, $\bm{W}_{in}$ and $\bm{W}$, are randomly initialized once and are not trained. Only $\bm{W}_{out}$ is trained. The sparse matrices $\bm{W}_{in}$ and $\bm{W}$ are constructed to satisfy the Echo State Property \cite{Lukosevicius2009}. Following \cite{Pathak2018a}, $\bm{W}_{in}$ is generated such that each row of the matrix has only one randomly chosen nonzero element, which is independently taken from a uniform distribution in the interval $[-\sigma_{in}, \sigma_{in}]$. Matrix $\bm{W}$ is constructed with an average connectivity $\langle d \rangle$, and the non-zero elements are taken from a uniform distribution over the interval $[-1,1]$. All the coefficients of $\bm{W}$ are then multiplied by a constant coefficient for the largest absolute eigenvalue of $\bm{W}$, i.e. the spectral radius, to be equal to a value $\Lambda$, which is typically smaller than (or equal to) 1.
To train the PI-ESN, hence $\bm{W}_{out}$, a combination of the data available and the physical knowledge of the system is used: the components of $\bm{W}_{out}$ are computed such that they minimize the sum of (i) the error between the PI-ESN prediction and the measured system states, $E_d$, and (ii) the physical residual, $\mathcal{F}(\widehat{\bm{y}}(n))$, on the prediction of the ESN, $E_p$:
\begin{equation}
E_{tot} = \underbrace{\frac{1}{N_t}  \sum_{n=0}^{N_t-1} \frac{1}{N_z} \sum_{i=1}^{N_z} ||\widehat{z}_i (n) - z_i (n) ||^2 }_{E_d} + \underbrace{\frac{1}{N_t}  \sum_{n=0}^{N_t-1}  \frac{1}{N_y} \sum_{i=1}^{N_y} || \mathcal{F}(\widehat{y_i}(n))||^2 }_{E_p}
\label{eq:err_tot}
\end{equation}
where $||\cdot||$ is the Euclidean norm. 
The training of the PI-ESN for the reconstruction of hidden states is initialized as follow. Matrix $\bm{W}_{out}$ is split into two partitions $\bm{W}_{z,out}$ and $\bm{W}_{h,out}$, i.e. $\bm{W}_{out} = [\bm{W}_{z,out}; \bm{W}_{h,out}]$, which are responsible for the prediction of the observed states, $\widehat{\bm{z}}=\bm{W}_{z,out} [\bm{x}(n); \bm{u}(n); 1]$, and the hidden states, $\widehat{\bm{h}}=\bm{W}_{h,out}[\bm{x}(n); \bm{u}(n); 1]$,  respectively. $\bm{W}_{z,out}$ is initialized by Ridge regression of the data available for the measured states
\begin{align}
\bm{W}_{z,out} = \bm{Z} \bm{X}^T \left( \bm{X} \bm{X}^T + \gamma \bm{I}  \right)^{-1}
\end{align}
where $\bm{Z}$ and $\bm{X}$ are respectively the horizontal concatenation of the measured states, $\bm{z} (n)$, and associated ESN states, inputs signals and biases, $[\bm{x}(n);\bm{u}(n);1]$ at the different time instants during training; $\gamma$ is the Tikhonov regularization factor \cite{Lukosevicius2009}; and $\bm{I}$ is the identity matrix. Matrix $\bm{W}_{h,out}$ is randomly initialized to provide an initial guess for the optimization of $\bm{W}_{out}$. The optimization process modifies the components of $\bm{W}_{out}$ to obtain  the hidden states, while ensuring that the predictions on the hidden states satisfy the physical equations. The optimization is performed with a stochastic gradient method (the Adam-optimizer \cite{Kingma2015}) with a learning rate of 0.0001.

\section{Results and Discussions}
\label{sec:results}
The approach described in Sect.~\ref{sec:method} is tested for the reconstruction of the chaotic Lorenz system, which is described by \cite{Lorenz1963}:
\begin{equation}
    \dot{\phi_1} = \sigma (\phi_2-\phi_1), \hspace*{11pt}
    \dot{\phi_2} = \phi_1 (\rho-\phi_3)-\phi_2, \hspace*{11pt}
    \dot{\phi_3} = \phi_1 \phi_2-\beta \phi_3 \label{eq:Lorenz}
\end{equation}
where $\rho=28$, $\sigma = 10$ and $\beta=8/3$. The size of the training dataset is $N_t=20000$ with a timestep between two time instants of $\Delta t = 0.01$. An explicit Euler scheme is used to obtain this dataset. We assume that only measurements of $\phi_1$ and $\phi_2$ are available for the training of the PI-ESN and the state $\phi_3$ is to be reconstructed. The parameters of the reservoir of the PI-ESN are taken to be: $\sigma_{in} = 1.0$, $\Lambda = 1.0$ and $\langle d \rangle = 20$. For the initialization of $\bm{W}_{z,out}$ via Ridge regression, a value of $\gamma = 10^{-6}$ is used for the Tikhonov regularization. These values of the hyperparameters are taken from previous studies \cite{Lu2017}, who performed a grid search. 

\subsection{Reconstruction of hidden states}
\begin{figure}[!ht]
    \centering
    \includegraphics[width=0.60\textwidth]{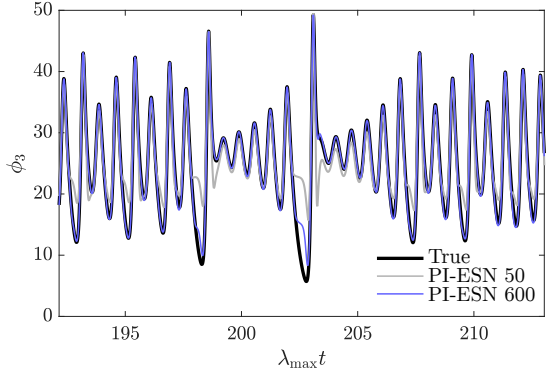}
    \caption{Reconstruction of $\phi_3$.}
    \label{fig:Lorenz_recons}
\end{figure}

In Fig. \ref{fig:Lorenz_recons} where the time is normalized by the largest Lyapunov exponent, $\lambda_{\max}=0.934$, the reconstructed $\phi_3$ time series is shown for the last 10\% of the training data for PI-ESNs with reservoirs of 50  and 600 units. (The dominant Lyapunov exponent is the exponential divergence rate of two system trajectories,  which are initially infinitesimally close to each other.) The small PI-ESN (50 units) can satisfactorily reconstruct the hidden state, $\phi_3$. The accuracy slightly deteriorates when $\phi_3$ has very large minima or maxima (e.g., $\lambda_{\max}t=202$). However, the large PI-ESN (600 units) shows an improved accuracy. The ability of the PI-ESN to reconstruct $\phi_3$, which is not present in the training data, is a key-result. 
The reconstruction is enabled exclusively by the knowledge of the physical equation, which is constrained into the training of the PI-ESN. This constraint allows the PI-ESN to deduce the evolution of $\phi_3$  from $\phi_1$ and $\phi_2$. Conversely, with neither the physical equation nor \textit{training data for $\phi_3$}, a data-only ESN cannot learn and reconstruct $\phi_3$ because it has no information on it.
\subsection{Effect of noise}
\begin{figure}[!ht]
    \centering
    \includegraphics[width=0.95\textwidth]{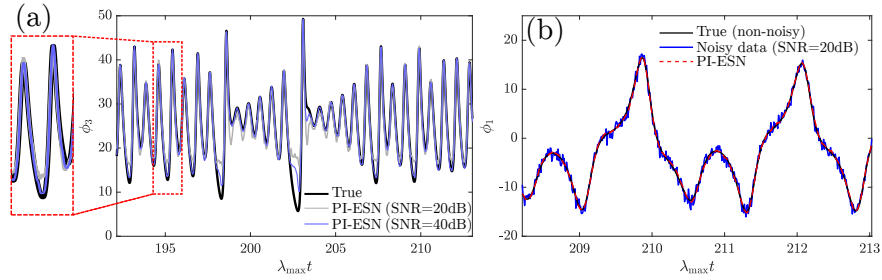}
    \caption{(a) Reconstruction of $\phi_3$ with PI-ESN of 600 units trained from noisy data (with zoomed inset). (b) Prediction of $\phi_1$.}
    \label{fig:Lorenz_recons_noise}
\end{figure}

\begin{figure}[!ht]
    \centering
    \includegraphics[width=0.60\textwidth]{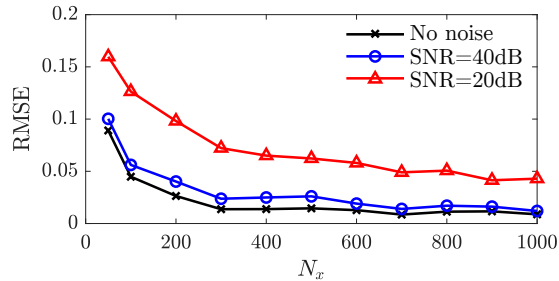}
    \caption{RMSE of the reconstructed $\phi_3$ time series in the training data.}
    \label{fig:Lorenz_recons_RMSE}
\end{figure}

As the ultimate objective is to work with real-world experimental data, the effect of noise on the results is investigated. The training data for $\phi_1$ and $\phi_2$ are modified by adding Gaussian noise to the original signal to imitate additive measurements noise. Two Signal-to-Noise Ratios (SNRs) of 20 dB and 40 dB are considered. The results of the reconstructed $\phi_3$ time series from the PI-ESN trained with the noisy training data are presented in Fig. \ref{fig:Lorenz_recons_noise}. Despite the presence of noise in the training data, the PI-ESN well reconstructs the non-noisy $\phi_3$ signal. This means that the physical constraints in Eq. (\ref{eq:err_tot}) act as a physics-based smoother (or denoiser) of the noisy data. This can be appreciated also in the {\it prediction} of measured states. Figure \ref{fig:Lorenz_recons_noise}b shows the prediction of state $\phi_1$: the non-noisy original data (full black line) and the prediction from the PI-ESN (dashed red line) overlap. This means that the PI-ESN provides a denoised prediction after training.
Finally, Fig. \ref{fig:Lorenz_recons_RMSE} shows the root mean squared error of the reconstructed hidden state $\widehat{\phi_3}$, $RMSE = \sqrt{ \frac{1}{N_t} \sum_{n=0}^{N_t-1}(\phi_3(n) - \widehat{\phi_3}(n))^2 }$, for PI-ESNs of different reservoir sizes and noise levels, 
where $\phi_3(n)$ is the reference non-noisy data, which we wish to recover. For the non-noisy case,  there is a large decrease in the RMSE when the PI-ESNs has 300 units (or more). With noise, the performance between the non-noisy and low-noise ($\textrm{SNR}=40$ dB) cases are similar, whereas for a larger noise level ($\textrm{SNR}=20$ dB), a larger reservoir is required to keep the RMSE small, as it may be expected. This suggests that the PI-ESN approach may be robust with respect to noise. 

\section{Conclusions and future directions}
\label{sec:conclusion}
We extend the Physics-Informed Echo State Network to reconstruct the hidden states in a chaotic dynamical system. The approach combines the knowledge of the system's physical equations and a small dataset.
It is shown, on a prototypical chaotic system, that this method can 
(i) accurately reconstruct the hidden states;
(ii)  accurately reconstruct the states with training data contaminated by noise; 
and 
(iii) provide a physics-based smoothing of the noisy measured data. 
Compared to other reconstruction approaches, the proposed framework does not require any data of the hidden states during training. This has the potential to enable the reconstruction of unmeasured quantities in experiments of higher dimensional chaotic systems, such as fluids. This is being explored in on-going studies. Future work also aims at assessing the effect of imperfect physical knowledge on the  reconstruction of the hidden states.

This paper opens up new possibilities for the reconstruction and prediction of unsteady dynamics from partial and noisy measurements.
%
\bibliographystyle{splncs04}
\bibliography{library_iccs2020_v2}

\end{document}